# Faddeev-Senjanovic Quantization of SU(n) N=2 Supersymmetric Gauge Field System with Non-Abelian Chern-Simons Topological Term and Its Fractional Spin

Yong-Chang Huang[1,2,*]   Qiu-Hong Huo[1]

[1] Institute of Theoretical Physics, Beijing University of Technology, Beijing 100022, China

[2] CCAST (World Laboratory), Beijing 100080, China

**Abstract**

Using Faddeev-Senjanovic path integral quantization for constrained Hamilton system，we quantize SU(n) N=2 supersymmetric gauge field system with non-abelian Chern-Simons topological term in 2+1 dimensions, and use consistency of a gauge condition naturally to deduce another gauge condition. Further, we get the generating functional of Green function in phase space, deduce the angular momentum based on the global canonical Noether theorem at quantum level, obtain the fractional spin of this supersymmetric system, and show that the total angular momentum has the orbital angular momentum and spin angular momentum of the non-abelian gauge field. Finally, we find out the anomalous fractional spin and discover that the fractional spin has the contributions of both the group superscript components and $A_0^s(x)$ charge.

# 1. Introduction

---

[*] Email address: ychuang@bjut.edu.cn

Supersymmetric Chern-Simons systems have been investigated in some references[1,2,3]. Attention has been given to a connection between extended supresymmetry and the existence of self-dual solutions[2]. A SU(n) N=2 supersymmetric gauge field model is constructed[4], which gives a system with non-topological self-dual solutions. Fractional spin and statistics have important meanings in explaining the quantum Hall effects [5,6,7] and high-$T_c$ superconductivity phenomena[8]. Fractional spin may appear in gauge theories with Chern-Simons(CS) topological term. CS gauge field does not have its own real dynamics, its dynamics comes from the fields to which it is coupled[9,10,11,12]. It is interesting to study the supersymmetric anyon system, because both spinor fields and scalar fields are naturally contained in supersymmetric fields. It is also a natural way to treat fractional spin and statistics by means of the supersymmetric model[4]. Many authors studied the angular momentum of Chern-Simons system by energy-momentum tensor and classical Noether theorem[9,10,11,12,13,14], and obtained the fractional spin character. But the conclusions are deserved to be discussed at quantum level by the phase-space path integral method, because the phase-space path integral method is more fundamental than the configuration-space path integral method.

It is the purpose of this paper to study the property of fractional spin of the SU(n) N=2 supersymmetric gauge field system with non-Abelian Chern-Simons topological term at the quantum field level. In Section 2, we introduce the supersymmetric gauge field system with non-Abelian Chern-Simons topological term and analyze the constraint structure; in Section 3, this model is quantized according to the rule of Faddeev-Senjanovic path integral quantization, and we get the generating functional of Green function in phase space; in section 4, based on global canonical Norther theorem in the path integral form, we deduce the angular momentum and present the fractional spin of this system at the quantum field level, we also compare the result with the model without the gauge fields term; in the last section, we make summary and conclusion.

## 2. Supersymmetric Gauge Field System with Non-Abelian Chern-Simons topological term and its constraint analysis

In Ref.[15], A SU(n) $N = 2$ Supersymmetric Gauge Field System with Non-Abelian Chern-Simons topological term in 2+1 dimensions was constructed. Using Wess-Zumino gauge, the action is expressed in terms of component fields as

$$S = \int d^3x [-\frac{1}{4}G^{ab,r}G^r_{ab} + \frac{1}{2}\kappa\varepsilon^{abc}(A^r_a\partial_b A^r_c + \frac{1}{3}f^{rsu}A^r_a A^s_b A^u_c) + i\psi^{+\alpha}(\gamma^a)^\rho_\alpha D_a\psi_\rho$$
$$+ (D^a\varphi)^+ D_a\varphi + \frac{1}{2}i\lambda^\alpha(\gamma^a)^\rho_\alpha D_a\lambda_\rho + \frac{1}{2}i\chi^\alpha(\gamma^a)^\rho_\alpha D_a\chi_\rho + \frac{\kappa}{2}\lambda^{\alpha,r}\lambda^r_\alpha$$
$$+ \frac{\kappa}{2}\chi^{\alpha,r}\chi^r_\alpha + \frac{1}{2}(D^a N^r)D_a N_r - U(\varphi,\varphi^+,\psi,\psi^+,\lambda,\chi,N)], \qquad (1)$$

fixing the potential by requiring the conservation of the fermion-number, the potential term is [15]

$$U(\varphi,\varphi^+,\psi,\psi^+,\lambda,\chi,N) = f^{rsu}\chi^{\alpha,r}\lambda^s_\alpha N^u - i(\psi^{+\alpha}\lambda^r_\alpha T^r\varphi - \varphi^+ T^r\lambda^{\alpha,r}\psi_\alpha)$$
$$-\psi^{+\alpha}\chi^r_\alpha T^r\varphi - \varphi^+ T^r\chi^{\alpha,r}\psi_\alpha - \frac{n-1}{2kn}v^2\psi^{+\alpha}\psi_\alpha - \psi^{+\alpha}N^r T^r\psi_\alpha$$
$$+\varphi^+(N^r T^r + \frac{n-1}{2kn}v^2)(N^s T^s + \frac{n-1}{2kn}v^2)\varphi + \frac{1}{2}(\varphi^+ T^r\varphi + \kappa N^r)(\varphi^+ T^r\varphi + \kappa N^r), \qquad (2)$$

in Eq.(1), $D_a = \partial_a - iA^r_a T^r$ ($a = 1,2,3$), $G^r_{ab} = \partial_a A^r_b - \partial_b A^r_a + f^{rsu}A^s_a A^u_b$. $T^r$ are generators of SU(n) in fundamental representation and satisfy $[T^r, T^s] = if^{rsu}T^u$ and $tr(T^a T^b) = \delta^{ab}/2$, $v$ is expectation value of vacuum state. $\gamma$ matrices are $\gamma^0 = i\sigma^1$, $\gamma^1 = \sigma^2$, $\gamma^2 = i\sigma^3$, satisfying $\gamma^a\gamma^b = g^{ab} + i\varepsilon^{abc}\gamma_c$, the metric is $g_{ab} = diag(+1,-1,-1)$, and $\varepsilon^{012} = \varepsilon^{12} = 1$.

It can be seen that Lagrangian density (1) is singular in the sense of Dirac method. First we analyze the constraints of this system in phase space. The canonical momenta are defined as

$$\pi_\alpha = \frac{\partial_R \mathcal{L}}{\partial \dot{\phi}^\alpha}, \qquad (3)$$

where $\phi^\alpha$ stand for the component fields, the subscript "$R$" denotes the right derivative for $\phi^\alpha$. This definition is trivial for scalar and vector fields, but when it is applied to spinor fields, the Grassmann property must be considered. We also may omit the subscript "$R$" for convenience. The momenta conjugate to the component fields $A_a^r$, $\psi_\rho$, $\psi^{+\alpha}$, $\varphi$, $\varphi^+$, $\lambda_\rho, \chi_\rho, N^r$ respectively, are

$$\pi^{a,r} = \frac{\partial_R \mathcal{L}}{\partial \dot{A}_a^r} = -G^{0a,r} + \frac{1}{2}\kappa\varepsilon^{0ab} A_b^r, \tag{4.1}$$

$$\pi_{\psi_\rho} = \frac{\partial_R \mathcal{L}}{\partial \dot{\psi}_\rho} = i\psi^{+\alpha}\left(\gamma^0\right)_\alpha^\rho, \tag{4.2}$$

$$\pi_{\psi^{+\alpha}} = \frac{\partial_R \mathcal{L}}{\partial \dot{\psi}^{+\alpha}} = 0, \tag{4.3}$$

$$\pi_\varphi = \frac{\partial_R \mathcal{L}}{\partial \dot{\varphi}} = D_0 \varphi^+, \tag{4.4}$$

$$\pi_{\varphi^+} = \frac{\partial_R \mathcal{L}}{\partial \dot{\varphi}^+} = D_0 \varphi, \tag{4.5}$$

$$\pi_{\lambda_\rho} = \frac{\partial_R \mathcal{L}}{\partial \dot{\lambda}_\rho} = \frac{1}{2} i\lambda^\alpha \left(\gamma^0\right)_\alpha^\rho, \tag{4.6}$$

$$\pi_{\chi_\rho} = \frac{\partial_R \mathcal{L}}{\partial \dot{\chi}_\rho} = \frac{1}{2} i\chi^\alpha \left(\gamma^0\right)_\alpha^\rho, \tag{4.7}$$

$$\pi_{N^r} = \frac{\partial_R \mathcal{L}}{\partial \dot{N}^r} = D_0 N^r. \tag{4.8}$$

According to Dirac-Bergmann procedure[16], the primary constraints of the system should includes $\pi_0^r$, $\pi_{\psi_\rho}, \pi_{\psi^{+\alpha}}, \pi_{\lambda_\rho}, \pi_{\chi_\rho}$. The constraints referring to fermion fields have novel feature, and can be handled in a different procedure [17]. According to Dirac-Bergmann procedure, the primary constraints of the system are given by

$$\Gamma_1^r = \pi^{0,r} \approx 0, \tag{5.1}$$

$$\Gamma_2^\rho = \pi_{\psi_\rho} - i\psi^{+\alpha}\left(\gamma^0\right)_\alpha^\rho \approx 0, \tag{5.2}$$

$$\Gamma_3^\alpha = \pi_{\psi^{+\alpha}} \approx 0, \tag{5.3}$$

$$\Gamma_4^\rho = \pi_{\lambda_\rho} - \frac{1}{2}i\lambda^\alpha \left(\gamma^0\right)_\alpha^\rho \approx 0, \tag{5.4}$$

$$\Gamma_5^\rho = \pi_{\chi_\rho} - \frac{1}{2}i\chi^\alpha \left(\gamma^0\right)_\alpha^\rho \approx 0. \tag{5.5}$$

where symbol "$\approx$" means weak equality in Dirac sense [16]. The canonical Hamilton density corresponding to action (1) is given by

$$\begin{aligned}
\mathcal{H}_c &= \pi_\varphi \dot\varphi + \pi_{\varphi^+}\dot\varphi^+ + \pi^{a,r}\dot A_a^r + \pi_{\psi_\rho}\dot\psi_\rho + \dot\psi^{+\alpha}\pi_{\psi^{+\alpha}} + \pi_{\chi_\rho}\dot\chi_\rho + \pi_{\lambda_\alpha}\dot\lambda_\alpha + \pi_{N^r}\dot N^r - \mathcal{L} \\
&= \frac{1}{4}G^{ij,r}G_{ij}^r - \frac{1}{2}\kappa\varepsilon^{0ij}A_0^r\partial_i A_j^r - \frac{1}{2}\kappa\varepsilon^{0ij}A_i^r\pi_j^r - A_0^r\partial_i\pi^{i,r} - \frac{1}{2}\pi^{i,r}\pi_i^r \\
&\quad - f^{rsu}\pi^{i,r}A_0^s A_i^u - \frac{1}{8}\kappa^2 A_i^u A^{i,u} + iA_0^r\pi_{\psi_\rho}T^r\psi_\rho + \pi_{\varphi^+}\pi_\varphi + iA_0^r(\pi_\varphi T^r\varphi - \pi_{\varphi^+}T^r\varphi^+) \\
&\quad + iA_0^r\pi_{\lambda_\rho}T^r\lambda_\rho + iA_0^r\pi_{\chi_\rho}T^r\chi_\rho + \frac{1}{2}\pi_{N^r}\pi_{N^r} + iA_0^r\pi_{N^s}T^rN^s - i\psi^{+\alpha}(\gamma^j)_\alpha^\rho D_j\psi_\rho \\
&\quad - (D^j\varphi)^+ D_j\varphi - \frac{1}{2}i\lambda^\alpha(\gamma^j)_\alpha^\rho D_j\lambda_\rho - \frac{1}{2}i\chi^\alpha(\gamma^j)_\alpha^\rho D_j\chi_\rho - \frac{\kappa}{2}\lambda^{\alpha,r}\lambda_{\alpha,r} - \frac{\kappa}{2}\chi^{\alpha,r}\chi_{\alpha,r} \\
&\quad - \frac{1}{2}(D^jN^r)D_jN^r + U(\varphi,\varphi^+,\psi,\psi^+,\lambda,\chi,N).
\end{aligned} \tag{6}$$

Then, the total Hamiltonian is

$$H_T = \int_V d^2x(\mathcal{H}_c + \eta_1^r\Gamma_1^r + \eta_2^\rho\Gamma_2^\rho + \eta_3^\alpha\Gamma_3^\alpha + \eta_4^\alpha\Gamma_4^\alpha + \eta_5^\rho\Gamma_5^\rho), \tag{7}$$

where $\eta_1^r, \eta_2^\rho, \eta_3^\alpha, \eta_4^\alpha$, and $\eta_5^\rho$ are relative multipliers. The Possion bracket in this paper is defined as [18]

$$\{F(x),G(y)\}_{PB} = \int dz\left\{\frac{\delta_L F(x)}{\delta\phi^\alpha(z)}\frac{\delta_R G(y)}{\delta\pi_\alpha(z)} - (-1)^{n_F\cdot n_G}\frac{\delta_L G(y)}{\delta\phi^\alpha(z)}\frac{\delta_R F(x)}{\delta\pi_\alpha(z)}\right\}, \tag{8}$$

where $n_F$ and $n_G$ denote the Grassmann parities of $F(x)$ and $G(y)$, respectively. The consistency conditions $\dot\Gamma_1^r = \{\Gamma_1^r, H_T\}_{PB} \approx 0$ lead to secondary constraints

$$\Gamma_6^r = \partial_i\pi^{i,r} - f^{rsu}\pi^{i,s}A_i^u + \frac{1}{2}\kappa\varepsilon^{0lm}\partial_l A_m^r - i(\pi_\varphi T^r\varphi + \pi_{\varphi^+}T^r\varphi^+ + \pi_{\psi_\rho}T^r\psi_\rho$$

$$+ \pi_{\lambda_\rho}T^r\lambda_\rho + \pi_\chi T^r\chi + \pi_{N^s}T^rN^s) \approx 0. \tag{9}$$

While the consistencies $\dot\Gamma_2^r, \dot\Gamma_3^\alpha, \dot\Gamma_4^\alpha$ and $\dot\Gamma_5^\rho$ of the primary constraints lead to the equations for determining the Lagrange multipliers, then no further constraint occurs. In the following, we need to classify the constraints. The non-zero Possion brackets of all the constraints are

$$\{\Gamma_2^\alpha, \Gamma_3^\rho\}_{PB} = -i(\gamma^0)^{\alpha\rho}\delta(x-y), \tag{10.1}$$

$$\{\Gamma_2^r, \Gamma_6^\rho\}_{PB} = i\pi_{\psi_\rho} T^r \delta(x-y), \tag{10.2}$$

$$\{\Gamma_4^\alpha, \Gamma_4^\beta\}_{PB} = -i(\gamma^0)^{\alpha\beta}\delta(x-y), \tag{10.3}$$

$$\{\Gamma_5^\alpha, \Gamma_5^\beta\}_{PB} = -i(\gamma^0)^{\alpha\beta}\delta(x-y). \tag{10.4}$$

Constraint $\Gamma_1^r$ are the first class constraint, constraints $\Gamma_2^r$, $\Gamma_3^\alpha$, $\Gamma_4^\rho$, $\Gamma_5^\rho$ and $\Gamma_6^r$ are second class. We need to find the maximal set of first class constraints. By research, we find that constraint $\Gamma_6^r$ can be combined with $\Gamma_3^\alpha$ to get a first class constraint. Finally, the first class constraints are

$$\Lambda_1^r = \Gamma_1^r = \pi^{0,r} \approx 0 \tag{11.1}$$

$$\Lambda_2^r = -\pi_{\psi_\rho} T^r (\gamma^0)^\alpha_\rho \pi_{\psi^{\dagger\alpha}} + \partial_i \pi^{i,r} - f^{rsu}\pi^{i,s}A_i^u + \frac{1}{2}\kappa\varepsilon^{0lm}\partial_l A_m^r - i(\pi_\varphi T^r \varphi + \pi_{\varphi^+} T^r \varphi^+ + \pi_{\psi_\rho} T^r \psi_\rho$$
$$+ \pi_{\lambda_\rho} T^r \lambda_\rho + \pi_{\chi_\rho} T^r \chi_\rho + \pi_{N^s} T^r N^s) \approx 0 \ . \tag{11.2}$$

$\Lambda_1^r$, $\Lambda_2^r$ are also gauge transformation generators. In the meantime, the second-class constraints are

$$\theta_1^\rho = \Gamma_2^\rho = \pi_{\psi_\rho} - i\psi^{+\alpha}\left(\gamma^0\right)^\rho_\alpha \approx 0, \tag{12.1}$$

$$\theta_2^\alpha = \Gamma_3^\alpha = \pi_{\psi^{+\alpha}} \approx 0, \tag{12.2}$$

$$\theta_3^\rho = \Gamma_4^\rho = \pi_{\lambda_\rho} - \frac{1}{2}i\lambda^\alpha\left(\gamma^0\right)^\rho_\alpha \approx 0, \tag{12.3}$$

$$\theta_4^\rho = \Gamma_5^\rho = \pi_{\chi_\rho} - \frac{1}{2}i\chi^\alpha\left(\gamma^0\right)^\rho_\alpha \approx 0. \tag{12.4}$$

Therefore, we complete the classification of the constraints.

## 3 Faddeev-Senjanovic Path integral Quantization of the supersymmetric system

The further step is to choose two gauge-fixing conditions, which is essential for both canonical quantization and path integral quantization. We consider the Coulomb gauge

$$\Omega_1^r = \partial_i A_i^r \approx 0, \tag{13}$$

There is still gauge freedom in this system, because of the existence of two first-class constraints $\Lambda_1^r$ and $\Lambda_2^r$. Another gauge-fixing condition should be compatible with the Hamilton mechanism, one most natural manner is to choose the consistent condition

$$\Omega_2^r = \dot{\Omega}_1^r = \{\Omega_1^r, H_T\}_{PB} = \nabla^2 A_0^r - \partial_i \pi^{i,r} - f^{rsu}(\partial_i A_0^s) A_i^u = 0. \tag{14}$$

On the other hand, because general physical processes should satisfy quantitative causal relation [14,15], some changes ( cause ) of some quantities in (14) must lead to the relative some changes ( result ) of the other quantities in (14) so that (14)'s right side keeps no-loss-no-gain, i.e., zero, namely, (14) also satisfies the quantitative causal relation, which just makes the different quantities form a useful expression. And then we can obtain

$$\{\Lambda_1^r(x), \Omega_2^s(y)\}_{PB} = -\nabla^2 \delta^{rs} \delta^{(2)}(x-y), \tag{15.1}$$

$$\{\Lambda_2^r(x), \Omega_1^s(y)\}_{PB} = -\nabla^2 \delta^{rs} \delta^{(2)}(x-y). \tag{15.2}$$

According to Faddeev-Senjanovic quantization formulation, the phase space generating functional of Green function for this supersymmetric system is given by [22]

$$Z[0] = \int D\phi^\alpha D\pi_\alpha \prod_{i=1}^{2} \delta(\Lambda_i) \sum_{j=1}^{2} \delta(\theta_j) \sum_{k=1}^{2} \delta(\Omega_k) \det|\{\Lambda_i, \Omega_k\}| \left(\det|\{\theta_j, \theta_{j'}\}|\right)^{1/2}$$

$$\cdot \exp\{i \int d^3 x (\pi_\alpha \dot{\phi}^\alpha - \mathcal{H}_c)\}, \tag{16}$$

where

$$\phi^\alpha = \left(A_a^r, \varphi, \varphi^+, \psi, \psi^+, \lambda, \chi, N\right), \quad \pi_\alpha = \left(\pi^{a,r}, \pi_\psi, \pi_\varphi, \pi_{\varphi^+}, \pi_\lambda, \pi_\chi, \pi_N\right). \tag{17}$$

We separately calculate $|\{\Lambda_i, \Omega_k\}|, |\{\theta_j, \theta_{j'}\}|$ in equation (16). Taking use of (15), we obtain

$$|\{\Lambda_i, \Omega_k\}| = \left[\nabla^2 \delta^{rs} \delta^{(2)}(x-y)\right]^2. \tag{18}$$

Taking use of (10) and (12), we write out the $|\{\theta_j, \theta_{j'}\}|$

$$\left|\{\theta_j, \theta_{j'}\}\right| = \left[\delta^{\alpha\beta}\delta^{(2)}(x-y)\right]^4 . \tag{19}$$

Through (18) and (19), we find it interesting that both $\left|\{\Lambda_i, \Omega_k\}\right|$ and $\left|\{\theta_j, \theta_{j'}\}\right|$ are independent of field variables and can be ignored in the generating functional. Thus condition (14) coming from the consistent condition very naturally eliminates the gauge arbitrariness. Using the properties of the δ-function[18]

$$\delta(\Lambda) = \int \frac{\mathcal{D}\mu_l}{2\pi} \exp\left\{i\int d^3x \mu_l \Lambda\right\}, \tag{20}$$

we finally write out the phase space generating functional of Green function

$$Z[0] = \int D\phi^\alpha D\pi_\alpha D\lambda_i D\mu_j D\omega_k \exp\{i\int d^3x (\mathcal{L}_{eff}^p)\} , \tag{21}$$

where

$$\mathcal{L}_{eff}^p = \mathcal{L}^p + \lambda_i^r \Lambda_i^r + \omega_j^s \Omega_j^s + \mu_k^u \theta_k^u , \tag{22.1}$$

$$\mathcal{L}^p = \pi_\varphi \dot{\varphi} + \pi_{\varphi^+} \dot{\varphi}^+ + \pi^{a,r} \dot{A}_a^r + \pi_\psi \dot{\psi} + \dot{\psi}^+ \pi_{\psi^+} + \pi_\chi \dot{\chi} + \pi_\lambda \dot{\lambda} + \pi_N \dot{N} - \mathcal{H}_c . \tag{22.2}$$

Where $\lambda_i^r, \omega_j^s$, and $\mu_k^u$ are multipliers of first class constraints $\Lambda_i^r$, gauge fixing conditions $\Omega_j^s$, and second class constraints $\theta_k^u$, respectively.

## 4 Quantum angular momentum and Fractional Spin

Many articles discussed fractional spin character of abelian and non-abelian system by classical Noether Theorem. It is more meaningful to study the symmetry character at quantum level, especially in path integral form. First, we formulate the results of the quantal canonical Noether theorem[18]: If the effective action $I_{eff}^P = \int d^2x \mathcal{L}_{eff}^P$ is invariant in extended phase space under the following global transformation

$$x^{\mu'} = x^\mu + \Delta x^\mu = x^\mu + \varepsilon_\sigma \tau^{\mu\sigma}(x,\phi,\pi) , \tag{23.1}$$

$$\phi^{\alpha'}(x') = \phi^\alpha(x) + \Delta\phi^\alpha(x) = \phi^\alpha(x) + \varepsilon_\sigma \xi^{\alpha\sigma}(x,\phi,\pi) , \tag{23.2}$$

$$\pi'_\alpha(x') = \pi_\alpha(x) + \Delta\pi_\alpha(x) = \pi_\alpha(x) + \varepsilon_\sigma \eta_\alpha^\sigma(x,\phi,\pi) ,\qquad (23.3)$$

where $\varepsilon_\sigma$ are global infinitesimal arbitrary parameters $(\sigma = 1,2,\cdots,r)$, $\tau^{\mu\sigma}$、$\xi^\sigma$ and $\eta^\sigma$ are some smooth functions of canonical variables and space time, and if the Jacobian of the transformation (23) of the field variables is equal to unity, then, there are conserved laws at the quantum level

$$Q^\sigma = \int_V d^2 x[\pi(\xi^\sigma - \phi_{,k}\tau^{k\sigma}) - \mathcal{H}_{eff}\tau^{0\sigma}] = const ,\sigma = (1,2,\cdots,r) .\qquad (24)$$

We now deduce the angular momentum using the conserved quantities in $2+1$ dimensions. Consider the Lorentz transformation

$$\Delta x^\mu = \delta\omega^{\mu\upsilon} x_\upsilon,\qquad (25.1)$$

$$\Delta\phi^\alpha = \frac{1}{2}\delta\omega^{\mu\nu} \left(\Sigma^{\mu\nu}\right)^\alpha_\beta \phi^\beta,\qquad (25.2)$$

$$\Delta\pi_\beta = \frac{1}{2}\delta\omega^{\mu\nu} \left(\Sigma^{\mu\nu}\right)^\alpha_\beta \pi_\alpha.\qquad (25.3)$$

Under the spatial rotation in $x_i$ and $x_j$ plane, the effective canonical action $I_{eff} = \int d^3 x \mathcal{L}_{eff}^p$ is invariant, and the Jacobian of the spatial rotation transformation is equal to unity. We can write out the conserved angular momentum according to (24)

$$J = \int d^2 x \varepsilon^{0ij} \left[ x_i(\partial_j \phi^\alpha)\pi_\alpha + \frac{1}{2}\pi_\alpha \left(\Sigma_{ij}\right)^\alpha_\beta \phi^\beta \right]$$

$$= \int d^2 x \varepsilon^{0ij} [x_i \pi^{a,r}\partial_j A_a^r + \pi_i^r A_j^r + x_i \pi_{\psi_\rho}\partial_j \psi_\rho + x_i \pi_\varphi \partial_j \varphi + x_i \pi_{\varphi^+}\partial_j \varphi^+$$

$$+ x_i \pi_{\lambda_\alpha}\partial_j \lambda_\alpha + x_i \pi_{\chi_\rho}\partial_j \chi_\rho + x_i \pi_{N^s}\partial_j N^s + \frac{1}{2i}(\pi_\psi \gamma_i \gamma_j \psi + \pi_{\lambda_\alpha}\gamma_i \gamma_j \lambda_\alpha + \pi_{\chi_\rho}\gamma_i \gamma_j \chi_\rho)].\qquad (26)$$

The last term is related to spinor fields and is coincide with the result [9,10,11,12] obtained by classical Noether theorem. One can observe that the partial angular momentum given by non-abelian Chern-Simons topological term is

$$J_{cs}^N = \int d^2 x \varepsilon^{0ij} \left[ x_i \pi^{0,r}\partial_j A_0^r + x_i \pi^{k,r}\partial_j A_k^r + \pi_i^r A_j^r \right].\qquad (27)$$

Using (4.1) and (5.1), we express (27) as

$$J_{cs}^N = \int d^2 x \left[ \varepsilon^{ij} x_i \pi^{k,r}\partial_j A_k^r + \varepsilon^{ij}\pi_i^r A_j^r \right]$$

$$= \int d^2x \left[ -\varepsilon^{ij} x_i G^{0k,r} \partial_j A_k^r - \varepsilon^{ij} G_{oi}^r A_j^r + \frac{\kappa}{2} \varepsilon^{ij} \varepsilon^{kl} x_i A_l^r \partial_j A_k^r + \frac{\kappa}{2} \varepsilon^{ij} \varepsilon_{il} A^{l,r} A_j^r \right]. \tag{28}$$

The total angular momentum is written as

$$J = \int d^2x \varepsilon^{0ij} x_i [\pi_{\psi_\rho} \partial_j \psi_\rho + \pi_\varphi \partial_j \varphi + \pi_{\varphi^+} \partial_j \varphi^+ + \pi_{\lambda_\alpha} \partial_j \lambda_\alpha + \pi_{\chi_\rho} \partial_j \chi_\rho + \pi_{N^s} \partial_j N^s + G^{k0,r} \partial_j A_k^r]$$

$$+ \int d^2x \varepsilon^{0ij} [\frac{1}{2i}(\pi_{\psi_\rho} \gamma_i \gamma_j \psi_\rho + \pi_{\lambda_\alpha} \gamma_i \gamma_j \lambda_\alpha + \pi_{\chi_\rho} \gamma_i \gamma_j \chi_\rho) + G_{i0}^r A_j^r]$$

$$+ \frac{\kappa}{2} \int d^2x (\varepsilon^{ij} \varepsilon^{kl} x_i A_l^r \partial_j A_k^r + \varepsilon^{ij} \varepsilon_{il} A^{l,r} A_j^r)$$

$$= J_O + J_S + J_F \tag{29}$$

The first part $J_O$ stands for the orbital angular momentum, the second part $J_S$ expresses the spin angular momentum, the third part $J_F$ is proved to be related with the fractional spin angular momentum. (29) includes both fermion and boson parts. Using the properties $\varepsilon^{0ij} \varepsilon_{0jk} = -\varepsilon^{0ij} \varepsilon_{0kj} = -\delta_k^i$, we have

$$J_F = \frac{\kappa}{2} \int d^2x (\varepsilon^{ij} \varepsilon^{kl} x_i A_l^r \partial_j A_k^r + \varepsilon^{ij} \varepsilon_{il} A^{l,r} A_j^r)$$

$$= -\kappa \int d^2x (\varepsilon^{ij} x_i A_j^s \varepsilon^{lm} \partial_l A_m^s). \tag{30}$$

We consider the Eular-Lagrange equation corresponding to non-abelian Chern-Simons fields $A_a^r$, we obtain

$$G^{ab,s} f^{rsu} A_b^u - \partial_c G^{ac,r} + \kappa \varepsilon^{abc}(\partial_b A_c^r + \frac{1}{2} f^{rsu} A_b^s A_c^u) = J^{a,r}, \tag{31}$$

where

$$J^{a,r} = -\psi^{+\alpha}(\gamma^a)_\alpha^\rho T^r \psi - i(D^a \varphi^+) T^r \varphi + i\varphi^+ T^r (D^a \varphi) - \frac{1}{2} \lambda^\alpha (\gamma^a)_\alpha^\rho T^r \lambda_\rho$$

$$- \frac{1}{2} \chi^\alpha (\gamma^a)_\alpha^\rho T^r \chi_\rho + iT^r N^s D^a N^s \tag{32}$$

Letting $a = 0$ in (31), we have the equation

$$G^{0i,s} f^{rsu} A_i^u - \partial_i G^{0i,r} + \kappa \varepsilon^{0ij}(\partial_i A_j^r + \frac{1}{2} f^{rsu} A_i^s A_j^u) = J^{0,r}. \tag{33}$$

Using the defining equation (4.1), (33) is expressed as

$$-f^{rsu} \pi^{i,s} A_i^u + \partial_i \pi^{i,r} + \frac{1}{2} \kappa \varepsilon^{0ij} \partial_i A_j^r = J^{0,r}. \tag{34}$$

Considering the guage fixing condition (13) and (14), there is the relation on the hypersurface of constraints

$$\pi^{i,r} \approx \partial^i A_0^r. \tag{35}$$

We can obtain

$$\frac{1}{2}\kappa\varepsilon^{0ij}\partial_i A_j^r = J^{0,r} - \nabla^2 A_0^r = \left(J^{0,r}\right)'. \tag{36}$$

It can be checked that the following asymptotic form (37) [23] of the non-abelian vortex configuration satisfys (36), and is also compatible with the gauge fixing condition (13) and (14)

$$A_i^r = \frac{2}{\kappa}\varepsilon_{0ij}\partial_x^j \int d^2 y D(x,y)\left[J^{0,r}(y)\right]', \tag{37}$$

Where $D(x,y)$ is the Green function with the explicit form

$$D(x,y) = -\frac{1}{2\pi}\ln|x-y| + const. \tag{38}$$

Substituting (37) into (30), we find

$$J_F = -\kappa \int d^2 x(\varepsilon^{ij} x_i A_j^s \varepsilon^{lm}\partial_l A_m^s) = \frac{2Q^s Q^s}{\pi\kappa}, \tag{39}$$

where

$$Q^s = \int d^2 x \left[J^{0,s}(x)\right]' = \int d^2 x J^{0,s}(x) - \int d^2 x \nabla^2 A_0^s(x). \tag{40}$$

When taking $A_0^s(x) = J_{A_0}^s(x)\ln|x-x_0|$, we have

$$Q^s = -2\pi J_{A_0}^s(x_0) + \int d^2 x J^{0,s}(x). \tag{41}$$

This term (39) is the "anomalous one" which is interpreted as fractional spin. For consistent quantum mechanics, the coefficient $\kappa$ should be quantized so that $\kappa = m/4\pi$ with nonzero integer $m$ [24].

Contrary to the abelian case, the result (39) has the contribution of group component values. We can also find that, different with the non-Abelian Chern-Simons model without gauge field strength term, the conserved charge (40) includes the term of $A_0^s(x)$ charge. When $Q^s$ is replaced by the abelian charge $Q$ and the contribution of $A_0^s(x)$ charge is zero, this result is reduced to the common result [9,10,11,12,13,14].

If there is no gauge field strength term in the Lagrangian density (1), we also obtain the anomalous $J_F$, but the orbital angular momentum and spin angular momentum of the field $A_\mu^r$ will disappear, which can be seen from (29).

## 5. Summary and conclusion

Using the Faddeev-Senjanovic method of path integral quantization for the canonical constrained system, we quantize the $SU(n)$ N=2 supersymmetric non-abelian system with Chern-Simons topological term. First, we analyze the constraints in phase space. Then, we take the Coulomb gauge and use its consistency to deduce another gauge condition. According to Faddeev-Senjanovic quantization formulation, we obtain the phase space generating functional of Green function. Based on the global canonical Noether theorem, we deduce the angular momentum of this system and the partial angular momentum given by non-abelian Chern-Simons topological term. We find the partial angular momentum to be the "anomalous spin". We also find that the total angular momentum in this letter is different from the system without gauge field strength term, the results deduced from the system without gauge field strength term is missing the orbital angular momentum and spin angular momentum of the field $A_\mu^r$. Different from the abelian case and the non-Abelian case which does not contain gauge field strength term, we find that the conserved charge (39) has the contributions of the group superscript components and $A_0^s(x)$ charge. We also compare our method with Banerjee's method[9,10,11], in which Banerjee added a term proportional to the Gauss constraint to the Schwinger's energy-momentum tensor, and chose the multiplier in a covariant way. He also compared modified Schwinger's energy-momentum tensor and canonical angular momentum, and found the difference between two angular momenta to be a boundary term which can be interpreted as the fractional spin. In our method, the total angular momentum plays the role of the canonical angular momentum as in Ref. [9,10,11]. Furthermore, we systemically deduce the total angular momentum independent of any specific choice of ansatz, and find that the orbital angular momentum, spin angular momentum, and fractional spin angular momentum all appear in the total angular momentum.

**Acknowledgement:**

The authors are grateful for Prof. Z. P. Li for useful discussion.

The work is supported by National Natural Science Foundation of China (10435080) and Beijing Natural Science Foundation (Grant No. 1072005).